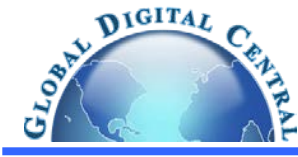

**Frontiers in Heat and Mass Transfer**

Available at www.ThermalFluidsCentral.org

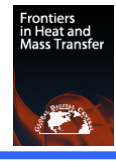

# SURROGATE-BASED OPTIMIZATION OF THERMAL DAMAGE TO LIVING BIOLOGICAL TISSUES BY LASER IRRADIATION

Nazia Afrin [a], Yuwen Zhang [b,*]

[a] *Department of Mechanical Engineering, St. Mary's University, San Antonio, TX 78228, USA*
[b] *Department of Mechanical and Aerospace Engineering, University of Missouri, Columbia, MO 65211, USA*

**ABSTRACT**

The surrogate-based analysis and optimization of thermal damage in living biological tissue by laser irradiation are discussed in this paper. Latin Hypercube Sampling (LHS) and Response Surface Model (RSM) are applied to study the surrogate-based optimization of thermal damage in tissue using a generalized dual-phase lag model. Response value of high temperature as a function of input variables and the relationship of maximum temperature and thermal damage as a function of input variables are investigated. Comparisons of SBO model and simulation results for different sample sizes are examined. The results show that every input variable individually has quadratic response to the maximum temperature and maximum thermal damage in highly absorbing tissues.

## 1. INTRODUCTION

Computational optimization is an important paradigm itself with a wide range of applications. In every sector in engineering, there are always needs to optimize something: computational time, cost, efficiency, energy consumption, maximize the profit, output, or performance. In many cases of engineering practice, optimization of objective functions comes from measurement of physical system or from computer simulations in a straightforward way. One reason behind this is that the simulation based objective function are often analytically interactable and another reason is the high computational cost of simulations. Long simulation times can be feasibly handled by using a surrogate model; which replaces the optimization of the original objective by an iterative re-optimization and updating of analytical computationally cheap surrogate models.

The unique characteristics of laser makes its application dramatically increased in every sector of science and engineering. In biomedical science, for example, laser is used in many treatments such as photodynamic therapy, cosmetic dermatology, laser mammography, and plastic surgery. Thermal damage is one of the major concerns of laser application in biomedical science. Pennes (1948) bioheat equation is the most widely used model to obtain temperature distribution in the living biological tissue. Welch's (1984) three-step model for laser induced thermal damage in biological tissue attracted many researchers' attentions. The thermal damage of the tissue was determined based on protein denaturation that was evaluated by a chemical rate process equation. Fourier's law, thermal wave model (Catteneo, 1958; Vernotte, 1958) were used to model the thermal effects on living biological tissues. Tzou's (1997) dual-phase lag DPL model introduced two different time delays between the temperature gradient and the heat flux which removed the precedence assumption of the thermal wave model. It allows either the temperature gradient precedes the heat flux or the heat flux to precede the temperature gradient in a transient process. The general bioheat equation can be express as

$$\rho c \frac{\partial T}{\partial t} = -\frac{\partial \mathbf{q}}{\partial x} + Q_L + Q_m + w_b \rho_b c_b (T_b - T) \qquad (1)$$

The main cause of the dual phase lag phenomena in the living biological tissue is the nonequilibrium between the blood and the surrounding tissue. Zhang (2009) derived a generalized DPL model based on nonequilibrium heat transfer in living biological tissue (Xuan and Roetzel, 1997). It was demonstrated that the phase lag times depended on intrinsic properties of blood and tissue, blood perfusion rate, and convection heat transfer. The values of phase lag times might vary from place to place in human body. The following equation with the tissue temperature as sole unknown was derived:

$$\tau_q \frac{\partial^2 T_s}{\partial t^2} + \frac{\partial T_s}{\partial t} = \alpha_{eff}[\nabla^2 T_s + \tau_T \frac{\partial}{\partial t}(\nabla^2 T_s)] + \frac{G}{(\rho c)_{eff}}(T_b - T_s) + \frac{(1-\varepsilon)Q_m + Q_L}{(\rho c)_{eff}} \qquad (2)$$

$$+ \frac{\varepsilon \rho_b c_b}{G(\rho c)_{eff}}[(1-\varepsilon)\frac{\partial Q_m}{\partial t} + \frac{\partial Q_L}{\partial t}]$$

Afrin *et al.* (2012) investigated temperature response and thermal damage induced by laser irradiation based on the non-equilibrium heat transfer in living biological tissues using a generalized DPL bioheat transfer model. It was shown that the generalized DPL model predicted significantly different temperature and thermal damage compared with classical DPL and Pennes bioheat model. They also studied the effects of laser parameters such as laser exposure time, laser irradiance, and coupling factor on thermal damage in living tissues. They found that the phase lag time for heat flux had more impact on the temperature earlier, while the phase lag time for temperature gradient had more impact on the temperature later. In another research, Afrin *et al.* (2017) studied the effects of uncertainties of laser exposure time, phase lag times, blood perfusion coefficient, scattering coefficient and diffuse reflectance of light on the thermal damage of living biological tissues by laser irradiation using a sample-based stochastic model. The variabilities of input and output materials were quantified using the coefficient of variance (COV) and interquartile range (IQR), respectively. The IQR analysis concluded that phase lag times for temperature gradient and heat flux, laser exposure time and blood perfusion rate had more significant influences on the maximum temperature and maximum thermal damage of the living tissues than the diffuse reflectance of light and scattering coefficient.

---

[*] *Corresponding Author. Email: zhangyu@missouri.edu*

SBO techniques are concerned with accelerating the optimization of simulation problems that are expensive and time consuming. In many engineering and scientific disciplines where complex numerical simulations or physical experiments need more data by additional experiments, surrogate modeling is relatively easier and cheaper to carry out (Bhosekar and Ierapetritou, 2018). It represents a class of optimization methodologies to rapidly identify the local and global optima with surrogate modeling. To build a surrogate model, Design of Experiments (DoE) methods are used to determine the location and distribution of the sample points in the design space. The goal is to gather maximum amount of information from a limited number of sample points. There are two categories of DoE methods in the literature: classical and modern DoE methods (Han and Zhang, 2012). Modern DoE methods, such as Latin hyperbolic sampling (LHS), orthogonal array design (OAD), and uniform design (UD), have great advantages for deterministic computer experiments without random error as arises in laboratory experiments. Kriging model and sometimes polynomial chaos expansion are combined with LHS are often used for surrogate modeling optimization (You *et al.*, 2009; Choi and Granshi, 2004). Uncertainty and sensitivity can also be analyzed with Monte Carlo analysis in conjunction with LHS (Helton and Davis, 2003).

In this paper, one of the most popular DoE for uniform sampling distribution, LHS, is used to choose random sampling in the design scheme. LHS is a method of sampling that can be used to produce input values for estimation of expectations of functions of output variable. The surrogate-based optimization is a feasible solution where the optimization of the high-fidelity model does not work or impractical (Koziel and Yang, 2011). The surrogate model optimization provides an approximation of the minimizer associated to the high-fidelity model. The surrogate-based optimization process can be summarized as follows:
1. Generate the initial surrogate model
2. Obtain an approximate solution using the surrogate optimization
3. Calculate the high-fidelity model at the approximate solution computed previously
4. Update the surrogate model using the new model data
5. Stop if the termination condition is satisfied; otherwise repeat step 2

The objective of this article is to investigate optimization of thermal damage in living biological tissue based on a generalized dual phase lag model. Surrogate-based optimization is used optimize thermal damage in highly absorbing biological tissue. In the following sections, the physical model and modeling with surrogate-based optimization are briefly summarize and results are discussed.

## 2. PHYSICAL MODEL

### 2.1 Governing equations

Figure 1 shows the physical model of the problem under consideration. With an initial temperature of T0 and a thickness of L, a finite slab of a biological tissue is considered in this study. At time t=0+, a flat-top laser beam is applied to the left surface of the slab. Since the spot size of the broad beam laser is much larger than the thickness of the thermally effected zone for the time period of interest, an 1D model can be considered to analyze the thermal response of the heated medium. The left surface depends on the laser light absorption and right boundary surface is thermally insulated (q=0). The generalized DPL model in terms of heat flux is used as a governing equation of this study as follows (Zhang, 2009):

$$\tau_q \frac{\partial^2 q}{\partial t^2} + \frac{\partial q}{\partial t} = \alpha_s \frac{\partial^2 q}{\partial x^2} + \alpha_s \tau_T \frac{\partial^3 q}{\partial t \partial x^2} - \alpha_s \frac{\partial Q_L}{\partial x} + \frac{G\alpha_s}{(1-\varepsilon)} \frac{\partial T}{\partial x} + \frac{G\alpha_s \tau_T}{(1-\varepsilon)} \frac{\partial^2 T}{\partial t \partial x} \quad (3)$$

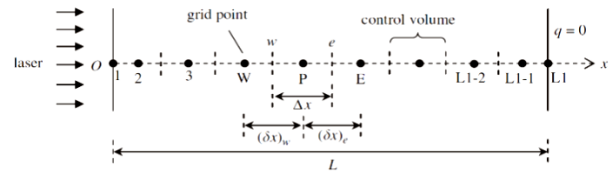

**Fig. 1** Physical model and grid system

When highly absorbed tissues are considered, the laser heating is approximated as a boundary condition of the second kind and the laser volumetric heat source or laser irradiance, $Q_L$, is zero. The boundary conditions are given below:

$$q = \phi_{in}(1-R_d) \quad \text{for } x = 0 \text{ when } 0 < t < \tau_L \quad (4)$$

$$q = 0 \quad \text{for } x = L \text{ when } 0 < t < \tau_L \quad (5)$$

where $\tau_L$ is the laser exposure time, $\phi_{in}$ is the incident laser irradiance, and $R_d$ is the diffuse reflectance of light at the irradiated surface. The initial condition is:

$$q = 0 \quad \text{for } 0 < x < L, t = 0 \quad (6)$$

The damage parameter is evaluated according to the Arrhenius equation [1]:

$$\Omega = A \int_{t_0}^{t_f} \exp(-\frac{E}{RT}) dt \quad (7)$$

where A is the frequency factor, $3.1 \times 10^{98}$ s$^{-1}$ (Welch, 1984); E is the activation energy of denaturation reaction, $6.28 \times 10^5$ J/mol (Welch, 1984); R is the universal gas constant, 8.314 J/ (mol. K); T is the absolute temperature of the tissue at the location where thermal damage is evaluated; $t_0$ is the time at onset of laser exposure; $t_f$ is the time of thermal damage evaluation. First performing integration of Eq. (3) over the control volume of grid point P and over the time step from t to t+Δt, then applying backward difference in time and piecewise-linear profile in space, the equation leads to the following form

$$a_P q_P^{t+\Delta t} = a_E q_E^{t+\Delta t} + a_W q_W^{t+\Delta t} + b \quad (8)$$

where

$$a_P = a_w + a_e + \frac{\tau_q \Delta x}{\Delta t} + \Delta x \quad (9)$$

$$a_E = \frac{\alpha_s \Delta t}{(\delta x)_e} + \frac{\alpha_s \tau_T}{(\delta x)_e} \quad (10)$$

$$a_W = \frac{\alpha_s \Delta t}{(\delta x)_w} + \frac{\alpha_s \tau_T}{(\delta x)_w} \quad (11)$$

$$b = [\frac{2\tau_q \Delta x}{\Delta t} + \Delta x + \frac{\alpha_s \tau_T}{(\delta x)_e} + \frac{\alpha_s \tau_T}{(\delta x)_w}] q_P^t - \frac{\alpha_s \tau_T}{(\delta x)_e} q_e^t - \frac{\alpha_s \tau_T}{(\delta x)_w} q_w^t - \frac{\tau_q \Delta x}{\Delta t} q_P^{t-\Delta t}$$
$$-\alpha_s \mu_a \Delta x \Delta t [\phi_{in}\{C_1 \frac{\partial}{\partial x} Exp[-k_1 x/\delta] - C_2 \frac{\partial}{\partial x} Exp[-k_2 x/\delta]\}]\Big|_P$$
$$+\frac{G\alpha_s}{(1-\varepsilon)} \frac{T_E^t - T_W^t}{2} \Delta t + \frac{G\alpha_s \tau_T}{(1-\varepsilon)} [\frac{T_E^{t+\Delta t} - T_W^{t+\Delta t}}{2} - \frac{T_E^t - T_W^t}{2}] \quad (12)$$

After replacing the values of the temperature-involved terms into Eq. (12) for the source term b, the discretization Eq. (8) becomes a linear system of algebraic equations that can be solved by TDMA (Tri-diagonal matrix algorithm). The temperature can be computed from the discretization form of the bioheat transfer model (Afrin and Zhang 2017).

## 2.2 Modeling with surrogate –based optimization

There are several surrogate models available in the literatures, such as response surface model (RSM), Kriging model, and radial basis function (RBFs) etc. (Barton, 1992; Simpson *et al.*, 1997; Jin and Simpson, 2001). The RSM consists of a group of mathematical and statistical techniques used in the development of an adequate functional relationship between a response of interest, y, and many associated inputs variables denoted by $x_1, x_2, ….x_k$. A relationship between them can be approximated by a degree polynomial model of the form

$$y = f'(x)\beta + \varepsilon \quad (13)$$

where $x = (x_1, x_2, ….. x_k)'$, f(x) is a vector function of p elements, β is a vector of p unknown constant coefficients, and ε is a random experimental error assumed to have a zero-mean value.

The LHS is a type of stratified Monte Carlo sampling. The basic idea is to make sample point distribution close to probability density function (PDF). The coding structure of LHS is given in Fig. 2. For producing a LHS of size N (McKay *et al.*, 1979), define $P=p_{jk}$ to a N×K matrix, where each column of P is an independent random permutation of (1,2…., N). Xjk is defined by

$$X_{jk} = F_k^{-1}(N^{-1}(p_{jk} - 1 + \xi_{jk})) \quad (14)$$

where $\xi_{jk}$ is the NK independent and identical distributed random variables independent of P (Stein, 1987). The nominal mean values of $\tau_L$, $w_b$, $\mu_s$, $R_d$, $\tau_T$, and $\tau_q$ are chosen as 5 s, 1.87×10-3 m$^3$ / (m$^3$ tissue s), 120 cm$^{-1}$, 0.05, 0.05 s, and 16 s, respectively. Standard deviation is taken to be 0.5 for all the input parameters. Defined MATLAB function is used to calculate LHS for the observed domain.

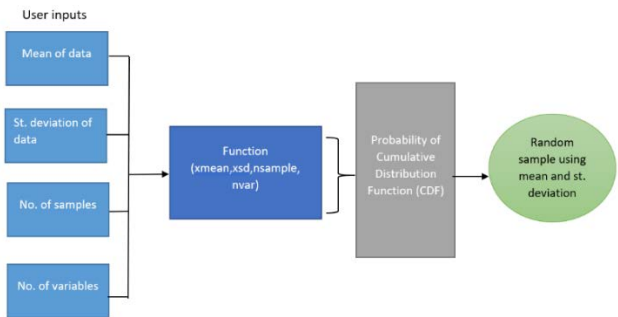

**Fig. 2** Coding structure of Latin Hyperbolic Sampling (LHS)

A series of simulation is run using a factorial design or fractional factorial design to identify the most significant explanatory variables. A focused design is used to determine the response to the explanatory variables within a range of interest. Using the least squares regression to a polynomial, RSM is produced to predict responses of values to the explanatory variables and parameters that are not simulated in the original model including an optimum (maximum or minimum) value. The inputs from LHS are $x=x_1, x_2, x_3…x_k$ and $S =[x(1), …., x(n)]^T$ and outputs $ys=[ ys (1), …., ys (n)]^T$. The true quadratic RSM can be written by Eq. (13). The pair (S, ys) denotes the sample data sets in the vector space. The second -degree model (d=2) can be defined as follows

$$y = \beta_0 + \beta_i x_i + \sum\sum\beta_{ij} x_i x_j + \sum\beta_{ij} x_{i2} + \varepsilon \quad (15)$$

where $\beta_0, \beta_i, \beta_{ii}$ and $\beta_{ij}$ are unknown coefficients to be determined. Next targets are first to establish a relationship between y and $x_1, x_2, x_3…x_k$, then determine optimum setting of $x_1, x_2, x_3…x_k$ that result in the maximum response over a certain region of interest. The least square estimator of β is

$$\beta = (U^T U)^{-1} U^T y_s \quad (16)$$

A response surface model can be generated using the least squares regression procedure. After determining all the coefficient of response surface, the approximated response can be obtained from untried point by Eq. (15). The model is specified for the problem by the following equation (Dixon, 2012):

$y = \beta_0 + \beta_1 x_1 + \beta_2 x_2 + \beta_3 x_3 + \beta_4 x_4 + \beta_5 x_5 + \beta_6 x_6 + \beta_{12} x_1 x_2 + \beta_{13} x_1 x_3 + \beta_{14} x_1 x_4 + \beta_{15} x_1 x_5 + \beta_{16} x_1 x_6 + \beta_{23} x_2 x_3 + \beta_{24} x_2 x_4 +$
$\beta_{25} x_2 x_5 + \beta_{26} x_2 x_6 + \beta_{34} x_3 x_4 + \beta_{35} x_3 x_5 + \beta_{36} x_3 x_6 + \beta_{45} x_4 x_5 + \beta_{46} x_4 x_6 + \beta_{56} x_5 x_6 + \beta_{11} x_1^2 + \beta_{22} x_2^2 + \beta_{33} x_3^2 + \beta_{44} x_4^2 + \beta_{55} x_5^2 + \beta_{66} x_6^2$ (17)

## 3. RESULTS AND DISCUSSIONS

Validation of generalized DPL model and surrogate based optimization (SBO) model is presented in Fig. 3. The SBO model is capable of predicting temperature within living biological tissues for highly absorbed tissue. It is shown from the Figure 3 that even though there are some deviation from the generalized DPL model, the SGO model is in good agreement with the generalized DPL model for highly absorbing tissue. The following properties of a living biological tissue are used for this analysis. Thermophysical properties of tissues: ρ=1000kg/m3, k=0.628W/(mK), c=4187J/(kg K); thermophysical properties of the blood vessel: $\rho_b$=1060kg/m$^3$, $c_b$=3860J/(kg K), $w_b$=1.87×10-3 m$^3$/(m$^3$ tissue s); optical properties: $\mu_s$=120.0cm-1, $\mu_a$=0.4cm-1, g=0.9; blood temperature: $T_b$=37ºC; and metabolic heat generation: $Q_m$=1.19×10$^3$ W/m$^3$. The thickness of the slab of tissue is L=5cm, and the initial temperature is $T_0$=37ºC. The diffuse reflectance $R_d$=0.05 is used for the laser light distribution of scattering tissue. The maximum relative error between the generalized DPL model and SBO model has found to be 9.75%.

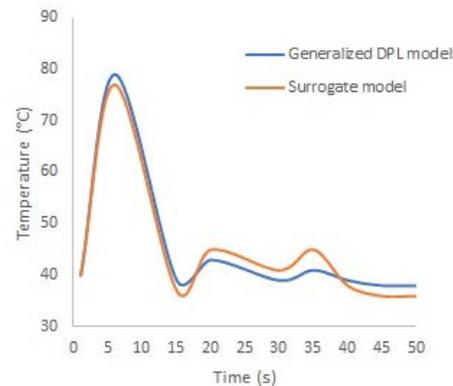

**Fig. 3** Validation of numerical simulation of tissue temperature for generalized DPL model and SBO method

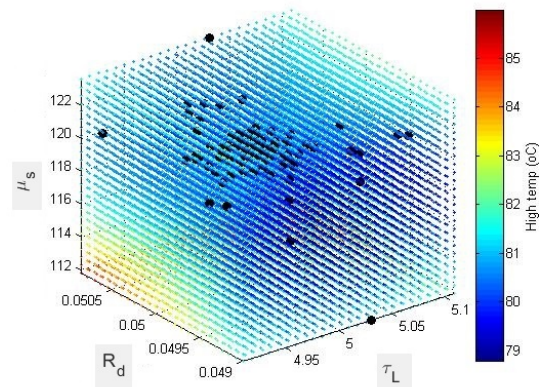

**Fig. 4** Response volume of high temperature as a function $\tau_L$, $R_d$, $\mu_s$

Response volume of maximum temperature as function of ($\tau_L$, $R_d$, $\mu_s$) and ($w_b$, $\tau_q$, $\tau_T$) are shown in Figs. 4 and 5, respectively. The figures show a general increase of temperature from the point (0,0,0) to the maximum values of the predictor variables ($\tau_L$, $R_d$, $\mu_s$, $w_b$, $\tau_q$, and $\tau_T$). Figure 4 shows the change of the maximum temperature in response of $\tau_L$, $R_d$ and $\mu_s$ which helps to assess the sensitivity of the maximum temperature in the living biological tissue.

In Fig. 5, the maximum temperature can assess the response to other three variables such as $w_b$, $\tau_q$ and $\tau_T$. Afrin *et al.* (2012) concluded that phase lag times for temperature gradient and heat flux, laser exposure time, and blood perfusion rate had more significant influences on the maximum temperature and maximum thermal damage of the living biological tissues. Therefore, the laser exposure time has chosen in this work to calculate the differences between the correct and estimated mean and variance.

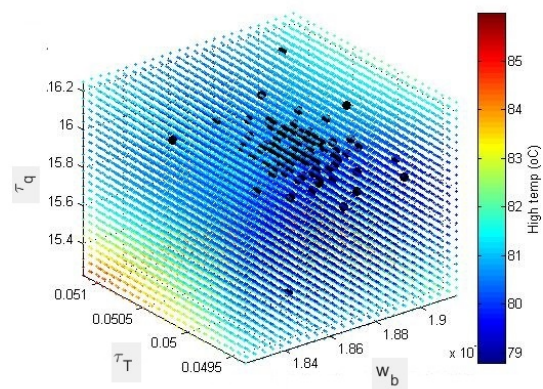

**Fig. 5** Response volume of high temperature as a function $w_b$, $\tau_q$, $\tau_T$

Table 1 shows the comparison of different of corrected and estimated mean of laser exposure time for SRS and LHS for different sample sizes (N). It can be seen from the table that with the increase of sample size (N), $\Delta\mu$ decreases significantly for LHS than SRS, which implies that LHS is better sampling procedure than SRS.

**Table 1.** Comparison of difference of corrected and estimated mean of laser exposure time for Simple Random Sampling (SRS) and Latin Hypercube Sampling (LHS) for different sample sizes (N)

| N (sample size) | $\Delta\mu$ (SRS) | $\Delta\mu$ (LHS) |
| --- | --- | --- |
| 50 | 0.167 | 0.056 |
| 100 | 0.325 | 0.078 |
| 500 | 0.237 | 0.0025 |

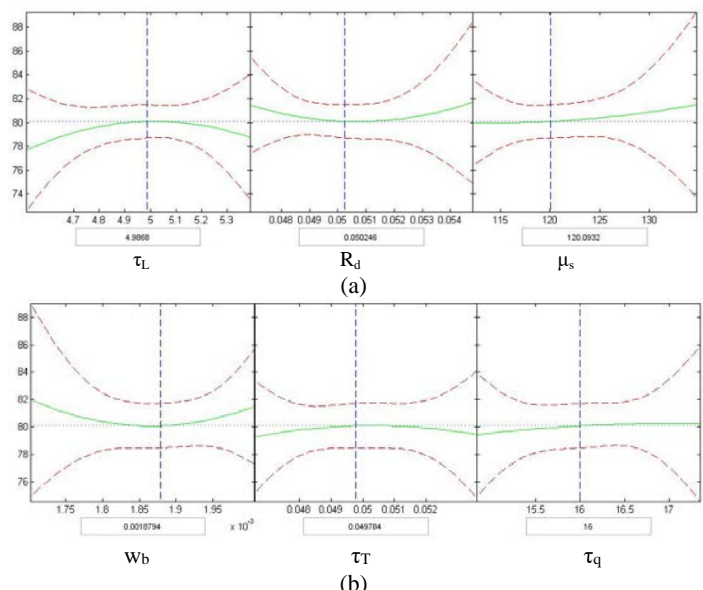

**Fig. 6** Maximum temperature as function of (a) $\tau_L$, $R_d$, $\mu_s$, and (b) $w_b$, $\tau_T$, and $\tau_q$

Figure 6 (a) and (b) show the maximum temperature as function of ($\tau_L$, $R_d$, $\mu_s$) and ($w_b$, $\tau_T$, and $\tau_q$), respectively. A sequence of plots is displayed where each of them showing contour of the surface response against a single predictor with all other predictors held constants. The plots in Fig. 6 (a) show a quadratic response of maximum temperature as a function of $\tau_L$, $R_d$ and linear increase of maximum temperature as a function of $\mu_s$. Figure 6 (b) shows a quadratic response of maximum temperature as a function of $w_b$ and linear increase as a function of $\tau_T$, and $\tau_q$. The two dashed curves represent 95% simultaneous confidence band for the fitted response. Predictor values of 5 inputs variables are shown on the horizontal axis and are marked by vertical dashed blue lines in the plots. The relationship of maximum temperature to each of the response variables two at a time instead of all are showing in Fig. 7. In this case, the value of $R_d$ at 0.05 and plotted the response of $\tau_L$ and $\mu_s$ (Fig. 7).

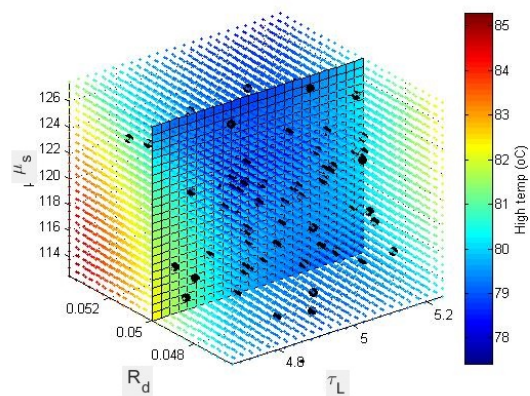

**Fig. 7** Maximum temperature as a function of $\tau_L$ and $\mu_s$ at the point $R_d=0.05$

Figure 8 show the response of the thermal damage as a function of (a) $\tau_L$, $R_d$, $\mu_s$, and (b) $w_b$, $\tau_T$, and $\tau_q$. The plots in Fig. 8 (a) show a quadratic response of thermal damage parameter as a function of $\mu_s$ and linear increase of damage parameter as a function of $R_d$ and $\tau_L$. Figure 8 (b) shows a quadratic response of damage parameter as a function of $w_b$, $\tau_T$ and $\tau_q$. The two dashed curves represent 95% simultaneous confidence band for the fitted response.

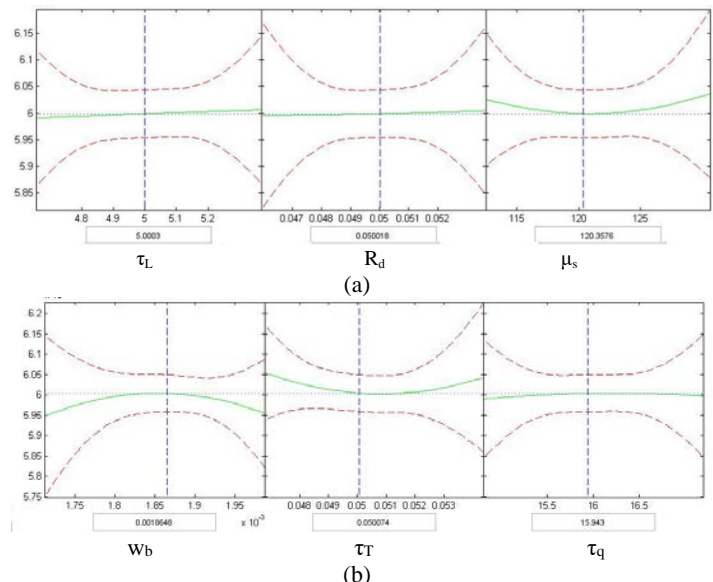

**Fig. 8** Damage parameter as function of (a) $\tau_L$, $R_d$, $\mu_s$, and (b) $w_b$, $\tau_T$, and $\tau_q$

The relationship of thermal damage to each of the response variables two at a time instead of all are showing in Fig. 9. In this case, the value of $R_d$ at 0.05 and plotted the response of $\tau_L$ and $\mu_s$.

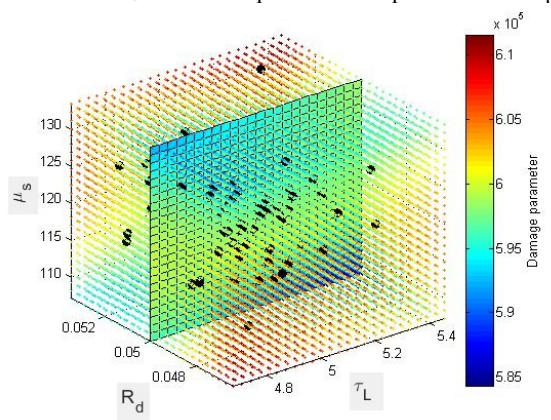

**Fig. 9** Damage parameter as a function of $\tau_L$ and $\mu_s$ at the point $R_d$=0.05

## 4. CONCLUSIONS

Surrogate based optimization has been used to optimize thermal damage in living biological tissues. Generalized DPL bioheat model based on the nonequilibrium heat transfer is applied to determine temperature response and thermal damage in highly absorbing tissues. Laser exposure time ($\tau_L$), blood perfusion ($w_b$), scattering coefficient ($\mu_s$), diffuse reflectance of light ($R_d$), phase lag times for temperature gradient ($\tau_T$), and for heat flux ($\tau_q$) are chosen as input variables. It is concluded that every input variables individually have quadratic response to the maximum temperature and maximum thermal damage in highly absorbing tissues. Comparison of difference between corrected and estimated mean of laser exposure time for Sample Random Sampling (SRS) and Latin Hypercube Sampling (LHS) for different sample size (N) shows that with the increase of sample size, $\Delta\mu$ decreases significantly for LHS than SRS.

## NOMENCLATURE

| | |
|---|---|
| $a$ | specific heat transfer area, m²/m³ |
| $c$ | specific heat, J/ (kg K) |
| $G$ | coupling factor between blood and tissue, W/ (m³ K) |
| $t$ | time, s |
| $T$ | average temperature, K |
| $q$ | heat flux vector, W/m² |
| $x$ | position vector, m |
| $w$ | blood perfusion rate, m³/m³ tissue |
| $Q_L$ | heat source due to hyperthermia therapy, W/m³ |
| $Q_m$ | source terms due to metabolic heating, W/m³ |
| $R_d$ | diffuse reflectance of light |
| $A$ | frequency factor, s⁻¹ |
| $R$ | universal gas constant, J/(mol K) |
| $E$ | energy of activation of denaturation reaction, J/mol |
| $\rho$ | density, kg/m³ |
| $\beta$ | vector of unknown constant coefficients |
| $\xi_{jk}$ | independent and identical distributed random variables |
| $\tau_q$ | phase lag time of the heat flux, s |
| $\tau_T$ | phase lag of the temperature gradient, s |
| $\tau_L$ | laser exposure time, s |
| $\alpha$ | thermal diffusivity, m²/s |
| $\varepsilon$ | porosity |
| $\varphi_{in}$ | incident laser irradiance |

**Subscripts**

| | |
|---|---|
| $b$ | blood vessel |
| $s$ | solid matrix (tissue) |